\documentclass{svproc}

\usepackage{mathtools}
\usepackage{amsmath}
\usepackage{amsfonts}
\usepackage{amssymb}
\usepackage{listings}
\usepackage{graphicx}
\usepackage{longtable}
\usepackage[parfill]{parskip}
\usepackage{bbm}
\usepackage{float}
\usepackage{adjustbox}
\usepackage{algorithm}
\usepackage{algorithmic}
\usepackage{caption}
\usepackage{subcaption}
\usepackage{hyperref}
\usepackage{multirow}
\usepackage{rotating}
\usepackage{newtxtext, makeidx, multicol, footmisc}
\usepackage[shortlabels]{enumitem}

\usepackage{float}

\title{Statistical network isomorphism}

\author{Pierre Miasnikof\inst{1} \thanks{Corresponding author: p.miasnikof@mail.utoronto.ca} \and
	Alexander Y. Shestopaloff\inst{2,3} \and
	Cristián Bravo\inst{4} \and
	Yuri Lawryshyn\inst{1} }
\authorrunning{P. Miasnikof et al.} 
\institute{
	University of Toronto, Toronto, ON, Canada, \\ \and
	Queen Mary University of London, London, United Kingdom, \\  \and
	Memorial University of Newfoundland, NL, Canada, \\ \and
	Western University, London, ON, Canada	
}

\date{}
\begin{document}
\maketitle

\begin{abstract}
Graph isomorphism is a problem for which there is no known polynomial-time solution. Nevertheless, assessing (dis)similarity between two or more networks is a key task in many areas, such as image recognition, biology, chemistry, computer and social networks. Moreover, questions of similarity are typically more general and their answers more widely applicable than the more restrictive isomorphism question. In this article, we offer a statistical answer to the following questions: a) {\it ``Are networks $G_1$ and $G_2$ similar?''}, b) {\it ``How different are the networks $G_1$ and $G_2$?''} and c) {\it ``Is $G_3$ more similar to $G_1$ or $G_2$?''}. Our comparisons begin with the transformation of each graph into an all-pairs distance matrix. Our node-node distance, Jaccard distance, has been shown to offer a good reflection of the graph's connectivity structure. We then model these distances as probability distributions. Finally, we use well-established statistical tools to gauge the (dis)similarities in terms of probability distribution (dis)similarity. This comparison procedure aims to detect (dis)similarities in connectivity structure, not in easily observable graph characteristics, such as degrees, edge counts or density.  We validate our hypothesis that graphs can be meaningfully summarized and compared via their node-node distance distributions, using several synthetic and real-world graphs. Empirical results demonstrate its validity and the accuracy of our comparison technique.
\end{abstract}

{\it {\bf Note on terminology:} For the sake of compactness, the work in this article focuses exclusively on simple graphs. We only consider unweighted, undirected graphs with no self-loops or multiple edges. Throughout this article, the terms graph and network are used interchangeably. Similarly, the terms vertex and node and the terms edge, arc, link and connection are used as synonyms.}

\section{Introduction}
Graph isomorphism is a problem for which there is no known polynomial-time solution. Nevertheless, assessing network (dis)similarity is a key task in many areas, such as image recognition, biology, chemistry, computer and social networks. In this article, we offer a statistical answer to the following questions: a) {\it ``Are networks $G_1$ and $G_2$ similar?''}, b) {\it ``How different are the networks $G_1$ and $G_2$?''} and c) {\it ``Is $G_3$ more similar to $G_1$ or $G_2$?''}. It is important to note here that these questions are more general and their answers more widely applicable than the more restrictive isomorphism question. 

We obtain these answers by first converting networks (graphs) into an all pairs distances matrix. To achieve this transformation, we use Jaccard distance instead of the typically used shortest-path or the also common random walk-based distances (e.g., commute, resistance,...). Previous work has highlighted the shortcomings of shortest-path \cite{RDist2021} and random walk-based distances \cite{vonLuxNIPS10,vonLux14,medoidsPPS19}. The advantages of Jaccard distance, especially its relation to connectivity structure, have also been demonstrated \cite{Camby17,PMCplxNets2020,PMCplxNets2022}.

Our comparison technique is focused on comparing each network's connectivity structure, not on easily observable graph characteristics such as vertex or edge counts. We argue that changes in connectivity may be indicative of critical network event occurrences, which makes structural conectivity-based (dis)similarity worthy of investigation. For example, the presence of denser subgraphs may indicate a loss of connection to the broader network and the appearance of bottlenecks, in a physical or computer network. They can also be an indicator of malicious activity, especially of the multi-party coordinated variety \cite{rla2008,rla2011,denseAnom2021}.  

As described later in this article, Jaccard distance also has a probabilistic interpretation. On the basis of this interpretation, we then compare networks as probability distributions of distances, using well-established statistical techniques. Our comparisons are not restricted to a few key statistical or graph characteristics, such as mean distance, mean degree or diameter. Instead, our conversion to a distance matrix and interpretation of these distances as a probability distribution captures each graph's entire connectivity structure.

\section{Previous work}
The comparison of static graphs and the study of temporal graphs are overlapping topics. Indeed, the study of temporal graphs naturally includes comparisons of snapshots of time-evolving graphs. In the past, several authors have highlighted the need to study graph similarity and their evolution over time. These authors have illustrated their claims using various areas of application, areas as varied as image recognition \cite{Bunke2000}, network robustness and resilience \cite{TangEtAl2013}, mobile telephony \cite{TangEtAl2011,TangEtAl2013,DuEtAl2018} and public transportation \cite{MudakoEtAl2019}. Notably, graph comparisons and temporal graphs remain current topics of inquiry \cite{HuangEtAl2020,CoupVreek2021,WangEtAl2022}.

A full review of the graph similarity and temporal graphs literature is beyond the scope of this short article. However, we wish to highlight the fact that this article is built upon the foundations of Schieber et al.~\cite{SchieberEtAl2017} and the very recent work of Wang et al.~\cite{WangEtAl2022}. These authors have modeled graphs as probability distributions. 

\section{Methods}
We model graphs as probability distributions of vertex-vertex distances. We too posit that graphs can be meaningfully summarized and compared on the basis of their node-node distances. Just as others before us, we begin by obtaining the distances between all vertex pairs. However, unlike in previous work, we use Jaccard distances \cite{JaccOrig,Camby17,PMCplxNets2020,PMCplxNets2022}. The main difference between earlier work and ours lies in the choice of node-node distance. 

Schieber et al.~\cite{SchieberEtAl2017} use shortest path distance. However, previous work has highlighted its shortcomings. For example, Akara-pipattana et al.~\cite{RDist2021} stated the following: {\it ``While intuitive and visual, this notion of distance is limited in that it does not fully capture the ease or difficulty of reaching point j from point i by navigating the graph edges. It does not say whether there is only one path of minimal length or many such paths, whether these paths can be straightforwardly located, or whether alternative paths are considerably or only slightly longer''}. In the past, Chebotarev and Shamis \cite{ChebotarevShamis2006a} as well as Fouss et al.~\cite{FoussEtAl2012} have also highlighted the unsuitability of shortest-path distance as a similarity measure between vertices. We have also echoed these assertions in recent publications and have demonstrated the superiority of the Jaccard distance as a reflection of graph structure \cite{PMCplxNets2020,PMCplxNets2022}.

In their very recent work, Wang et al.~\cite{WangEtAl2022} use a combination of embedding and Euclidean distance. While they report interesting results, this two-step process appears cumbersome and ill-suited to larger graphs, at first glance. Arguably, embedding graphs into vector space carries a non-trivial computational cost. In this specific case, the authors use the DeepWalk algorithm \cite{DeepWalk2014} to obtain their embedding. While the creators of DeepWalk claim their technique is scalable and parallelizable, it simulates random walks across the network. In contrast, Jaccard distance only relies on simple vertex-pair level arithmetic computations, instead of multiple layers of neighborhoods, and has been shown to offer an accurate reflection of graph structure \cite{PMCplxNets2020,PMCplxNets2022}. Its computation can also be easily performed incrementally or in parallel. In addition, several authors have highlighted the breakdown of the random-walk based commute (resistance) distance in the case of larger graphs \cite{vonLuxNIPS10,vonLux14,medoidsPPS19}.

We would also like to draw attention to the fact some authors restrict their comparisons to graphs with equal numbers of nodes \cite{GroheEtAl2018}. Yet, others are interested in the more general case of comparisons between graph with unequal numbers of nodes \cite{KoutraEtAl2011}. Because we compare connectivity through cumulative distributions, the number of nodes in each graph is not relevant. Our technique applies equally to either case.

\subsection{Vertex-vertex Jaccard distance}
The Jaccard distance separating two vertices $i$ and $j$ is defined as
\[
\zeta_{ij} = 1 - \underbrace{\frac{ \vert a_i \cap a_j \vert  }{ \vert a_i \cup a_j \vert}}_{s_{ij}} \in [0,1] \, .
\]
Here, $a_i \, (a_j)$ represents the set of all vertices with which vertex $i \, (j)$ shares an edge. The ratio $s_{ij}$ is the well known Jaccard similarity. The Jaccard distance ($\zeta_{ij}$) is its complement.

\subsubsection{Probabilistic interpretation of the Jaccard distance}
The Jaccard similarity ($s_{ij}$) between two nodes $i$ and $j$ can be interpreted probabilistically. Consider all nodes of a network, excluding $i$ and $j$, and select at random a node $k$. The Jaccard similarity is then an estimate of the (conditional) probability that both $i$ and $j$ are connected to $k$, given that at least one of $i$ and $j$ is connected to $k$. Mathematically, we express $s_{ij}$ as, 
\[
s_{ij} = P \left( \left( e_{ik} \land e_{jk} \right) \mid \left( e_{ik} \lor e_{jk)} \right) \right) \, ,
\] where $e_{ij}$ indicates the existence of an edge between nodes $i$ and $j$.

The Jaccard distance ($\zeta_{ij}$) is its complement. It can be interpreted as one of these two cases:
\begin{enumerate}[a)]
\item the (conditional) probability that $i$  is connected to $k$, but $j$ is not,
\end{enumerate}
(exclusive) or
\begin{enumerate}[b)]
\item  the (conditional) probability that $j$  is connected to $k$, but $i$ is not.
\end{enumerate}
Mathematically, we express it as
\[
\zeta_{ij} = 1 - P \left( \left( e_{ik} \land e_{jk} \right) \mid \left( e_{ik} \lor e_{jk)} \right) \right) = P \left( \left( e_{ik} \veebar e_{jk} \right) \mid \left( e_{ik} \lor e_{jk)} \right) \right) \, . 
\]

\subsubsection{From graph to empirical probability distribution} 
Once all distances $\zeta_{ij}$ have been obtained, we examine their statistical distribution. On the basis of the probabilistic interpretation of the $\zeta_{ij}$ just described, we treat these quantities as random variables. This model allows us to study and compare graphs as empirical probability distributions of node-node distances. 

\begin{figure}[]
	\centering
	\subfloat[ER]{ \includegraphics[width = 0.333\textwidth]{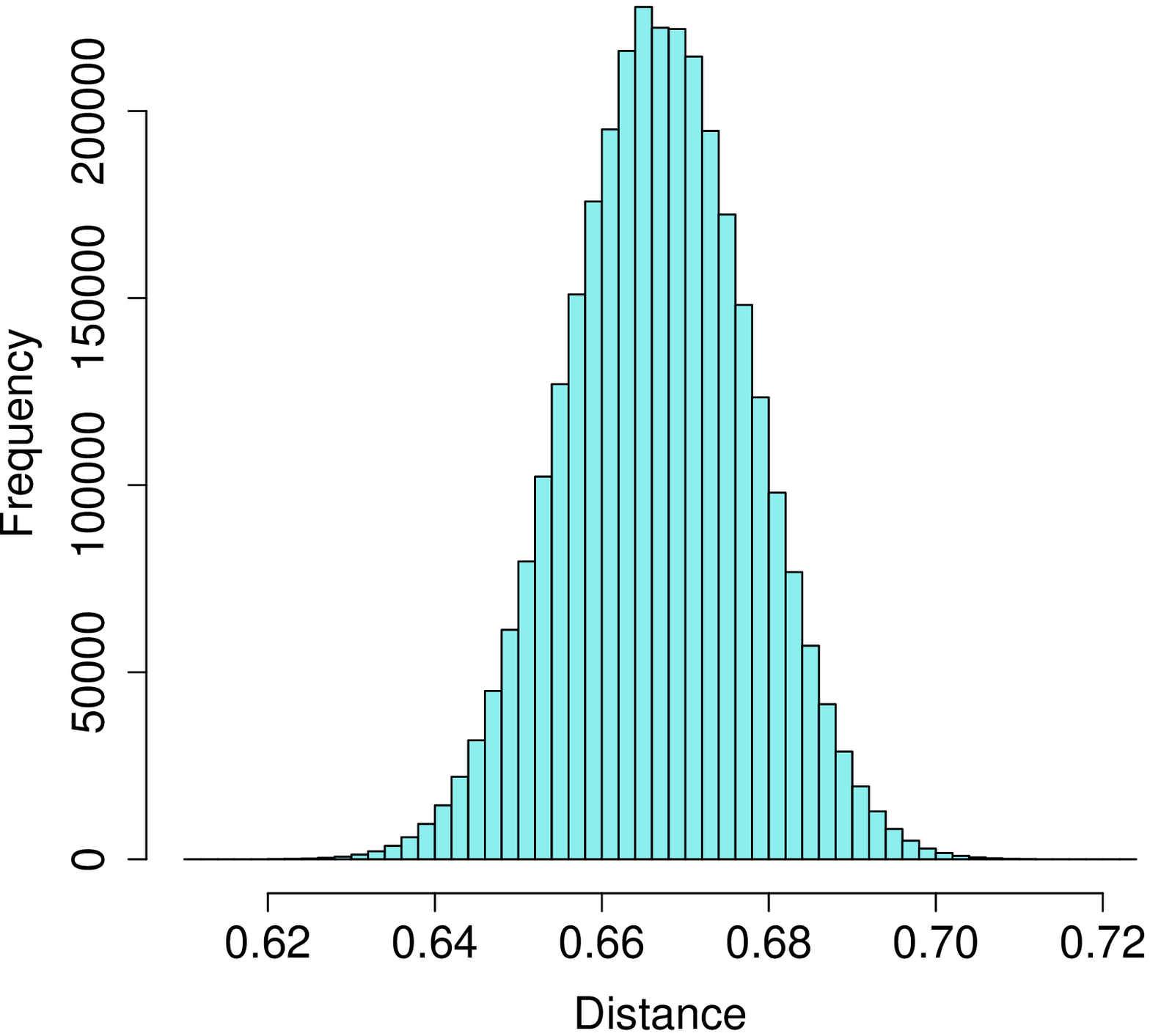} } 
	\subfloat[SBM]{ \includegraphics[width = 0.333\textwidth]{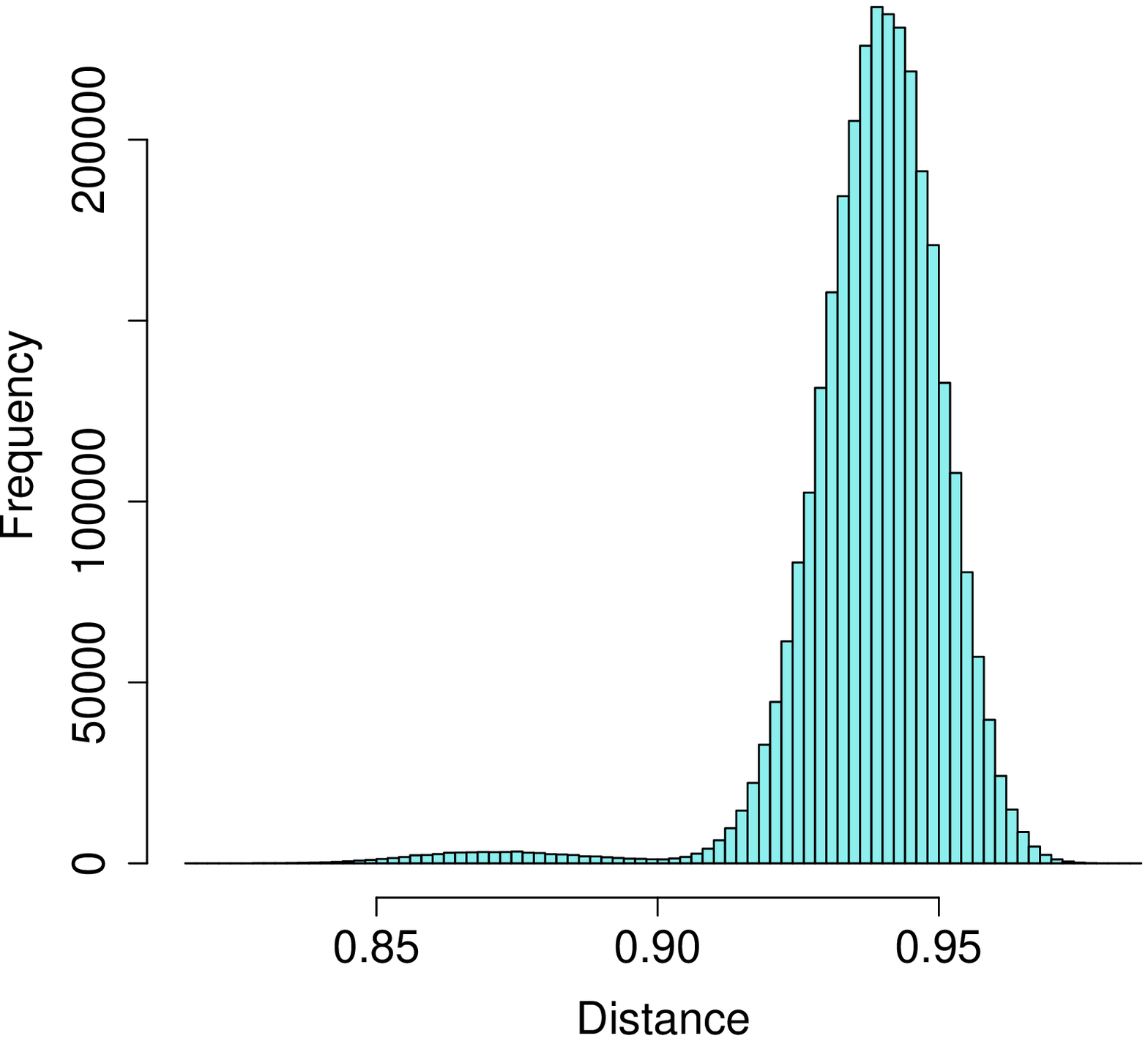} } 
	\caption{Distances as distributions}
	\label{distn}
\end{figure}

Figure~\ref{distn} illustrates the interpretation of a graph as a probability distribution. The image on the left shows the distribution of node-node distances for an  Erd\H{o}s-R\'enyi (ER) graph with an edge probability of $p=0.5$. The image on the right is of the distribution of distances between nodes of a stochastic block model graph (SBM) of varying cluster sizes and in/out edge probabilities of $0.9/0.1$.

The structural differences between these two graphs is immediately obvious. The ER graph's distances are symmetrically distributed about their mean, in a Gaussian-like pattern. In stark contrast, the SBM graph's distances are left-skewed and bi-modal. The left mode reflects distances between nodes in the same blocks, whereas the right mode reflects distances between nodes not in the same blocks. Naturally, this pattern does not occur under the ER model.

\subsection{Dissimilarity of probability distributions}
We compare the networks of interest via the empirical probability distributions of the Jaccard distances between their nodes. To perform these comparisons, we use the Kolmogorov-Smirnov (K-S) distance and the Wasserstein distance of order $p$. These distances are defined as follows: In a comparison between two networks, let $F_{1}(x)$ be the empirical cumulative distribution function (CDF) of Jaccard distances for the first network, and $F_{2}(x)$ the empirical CDF of Jaccard distances for the second network ($F^{-1}$ denotes the inverse CDF). The Kolmogorov-Smirnov (K-S) distance to compare $F_{1}$ and $F_{2}$ is defined as
\begin{equation*}
D = \sup_{x} \vert F_{1}(x) - F_{2}(x) \vert \quad (\in [0,1]) \, .
\end{equation*}
Meanwhile, the Wasserstein distance of order $p$ between $F_{1}$ and $F_{2}$ is defined as
\begin{equation*}
W_{p}(F_{1}, F_{2}) = \biggl( \int_{0}^{1} \vert F_{1}^{-1}(u) - F_{2}^{-1}(u) \vert^{p}du \biggr)^{1/p} \, .
\end{equation*} (In our experiments, we set the parameter $p=2$.)

The K-S distance metric $D$ is also a test statistic. In this specific case, it is a test statistic for the two-sample K-S test. The hypotheses of this test are listed below.
\begin{itemize}
	\item Null hypothesis ($H_o$) : the two samples are drawn from the same distribution
	\item Alternative hypothesis ($H_a$) : the two samples are drawn from different distributions
\end{itemize}
The $p$-values of the K-S test provide an interpretation and validation of the test statistic (distance). They give us the probability of obtaining a distance metric of the same or greater magnitude, under the (null) hypothesis that both samples were drawn from the same distribution. Small $p$-values provide evidence that the maximum vertical distance between the empirical CDFs of two compared graphs is statistically significantly different from zero. These $p$-values are obtained from the Kolmogorov distribution.

While the K-S distance is always contained in the interval $[0,1]$, the Wasserstein distance is not. To make comparisons more meaningful and easier to interpret, we transform the latter, so that it also lies on the same interval. After obtaining the quantity $W_p$, we perform the following transformation,
\[
\tilde{\mathcal{W}}_p = 1 - \exp(-W_p) \quad (\in [0,1]) \, .
\] In our comparisons, we use the quantity $\tilde{\mathcal{W}}_p$.

\section{Numerical results}
We validate our hypothesis that graphs can be meaningfully summarized and compared via their node-node distance distributions, using several synthetic and real-world graphs. Their key characteristics are reported in Table~\ref{graphchars}. The columns correspond to
\begin{itemize}
\item $\vert V \vert$: number of vertices,
\item $\vert E \vert$: number of edges,
\item $\mathcal{K}$: density,
\item $\min(D)$: minimum degree,
\item $\bar{D}$: mean degree,
\item $\max(D)$: maximumm degree and
\item $\vert CC \vert$: number of connected components.
\end{itemize}

\begin{table}[H]
	\centering
	\caption{Graph characteristics} \label{graphchars}
	\begin{tabular}{clccccccc}
		\hline
		\hline \\
		&            & $\vert V \vert$ & $\vert E \vert$ & $\mathcal{K}$  & $\min(D)$& $\bar{D}$ & $\max(D)$ & $\vert CC \vert$ \\
		\hline
		\multirow{7}{*}{{\begin{turn}{90} Synthetic \end{turn}}}  & ER.333     & 2,500           & 1,039,694       & 0.33 & 753        & 831.76             & 929        & 1  \\
		& ER.35      & 2,500           & 1,092,408       & 0.35 & 794        & 873.93      & 944        & 1  \\
		& ER.5       & 2,500           & 1,562,067       & 0.50 & 1,157      & 1,249.65    & 1,344      & 1  \\
		& ER.3332cc  & 2,500           & 887,948         & 0.28 & 45         & 710.36      & 843        & 2  \\
		& ER.333N1K  & 1,000           & 166,417         & 0.33 & 289        & 332.83      & 381        & 1  \\
		& SBM0701    & 2,495           & 348,674         & 0.11 & 230        & 279.50      & 334        & 1  \\
		& SBM0901    & 2,495           & 360,867         & 0.12 & 235        & 289.27      & 346        & 1  \\
		\hline
		\multirow{8}{*}{{\begin{turn}{90} Real-world \end{turn}}} & 1997/11/08 & 3,015           & 5,156           & 0.00 & 1          & 3.42        & 590        & 1  \\
		& 1997/11/09 & 3,011           & 5,150           & 0.00 & 1          & 3.42        & 589        & 1  \\
		& 1998/11/08 & 4,296           & 7,815           & 0.00 & 1          & 3.64        & 935        & 1  \\
		& 1998/11/09 & 4,301           & 7,838           & 0.00 & 1          & 3.64        & 938        & 1  \\
		& 1999/11/08 & 6,127           & 12,046          & 0.00 & 1          & 3.93        & 1,383      & 1  \\
		& 1999/11/09 & 3,962           & 7,931           & 0.00 & 1          & 4.00        & 837        & 1  \\
		& 2000/01/01 & 3,570           & 7,033           & 0.00 & 1          & 3.94        & 740        & 1 \\
		& 2000/01/02 & 6,474           & 12,572          & 0.00 & 1          & 3.88        & 1,458      & 1  \\
	 \\
		\hline
		\hline
	\end{tabular}
\end{table}

\begin{figure}[]
	\centering
	\subfloat[Synthetic]{ \includegraphics[width = 0.57\textwidth]{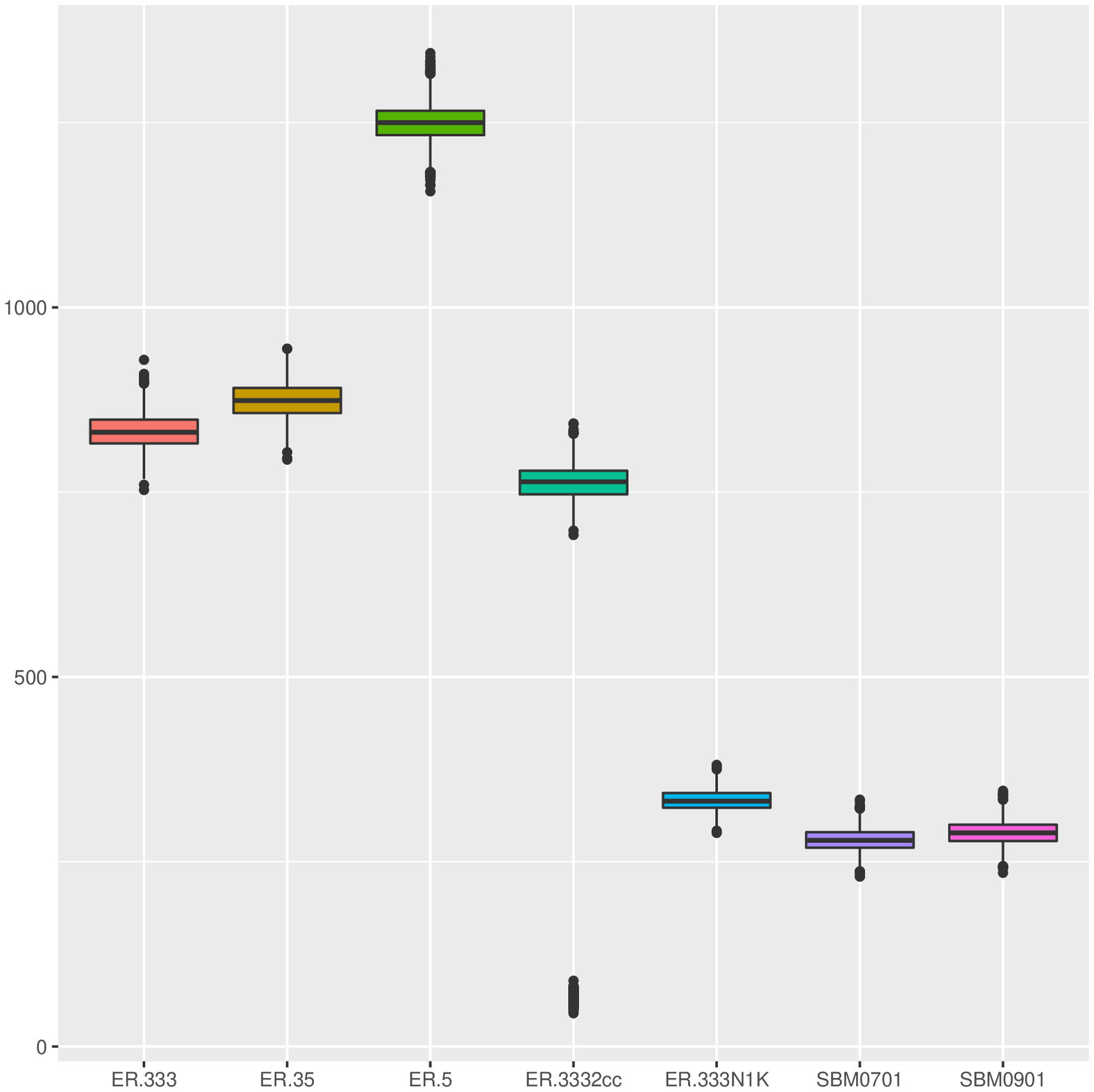} } 
	\subfloat[Real-world]{ \includegraphics[width = 0.57\textwidth]{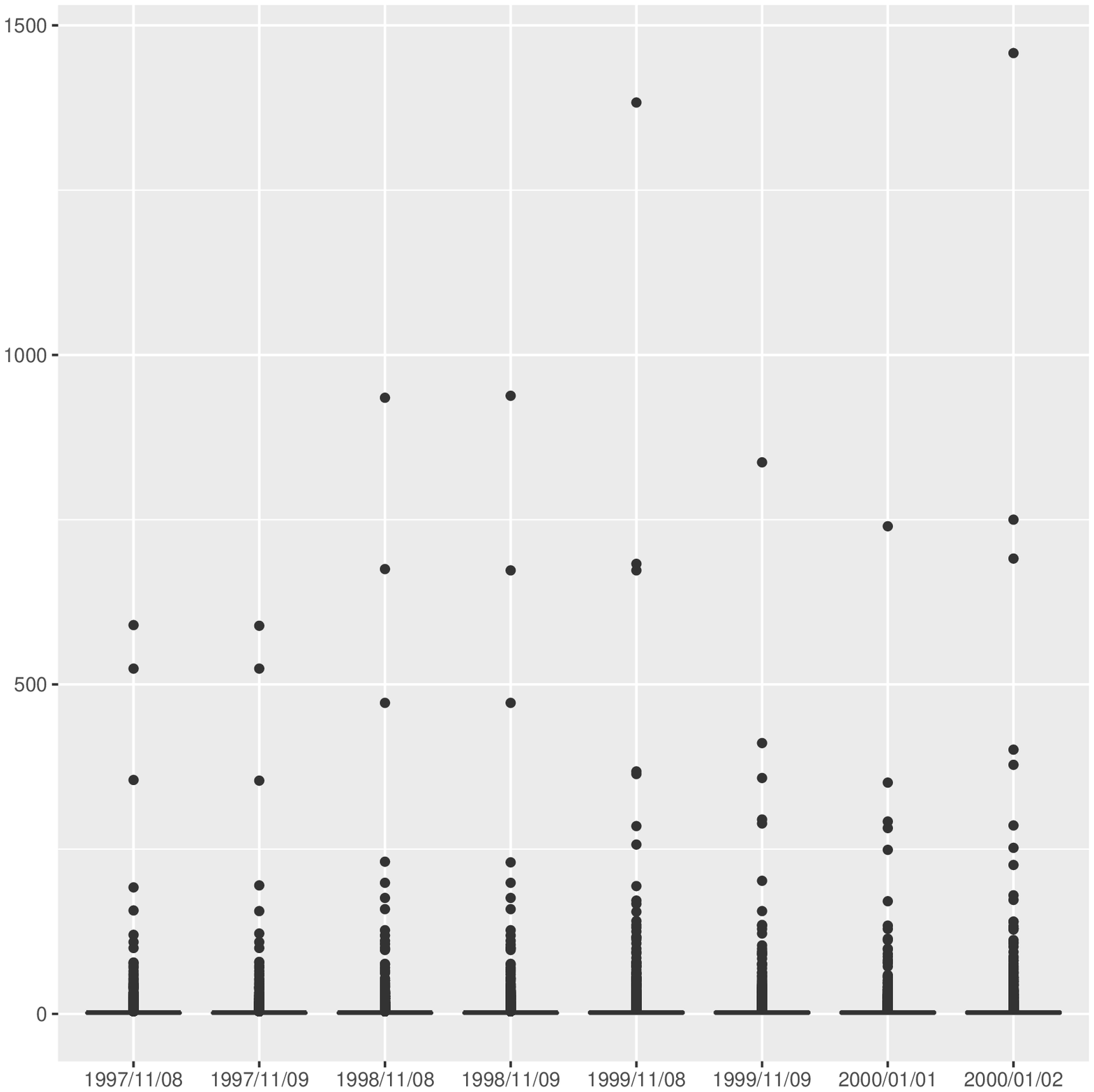} } 
	\caption{Degree distributions, box-plots}
	\label{degdistns}
\end{figure}

The generative models used to create the synthetic graphs are listed below. These graphs were generated using the NetworkX library \cite{NetworkX2008}.
\begin{itemize}
\item ER.333: ER with $n=2,500, \, p=0.333$
\item ER.35: ER with $n=2,500, \, p=0.35$
\item ER.5: ER with $n=2,500, \, p=0.5$
\item ER.3332cc: ER also with $p=0.333$, but with two connected components ($n_1= 2300, \, n_2 = 200$)
\item ER.333N1K: ER also with $p=0.333$, but with only $n = 1,000$ 
\item SBM0701: stochastic block model, with clusters in range of $[37,62]$ and $p_{in} = 0.7, \, p_{out} = 0.1$
\item SBM0901: stochastic block model, with clusters in range of $[37,62]$ and $p_{in} = 0.9, \, p_{out} = 0.1$
\end{itemize}
Meanwhile, the real-world graphs were obtained from the Harvard Dataverse repository \cite{RWGraphs2016}. These data sets are from the University of Oregon's ``Route Views Project''. Each graph contains a daily snapshot of a set of internet ``autonomous systems'' and their connections.

\begin{table}[]
	\centering
	\caption{Wasserstein distances, synthetic graphs} \label{Wd-synth}
	\begin{tabular}{|l|ccccccc|}
		\hline
		& ER.333 & ER.35 & ER.5 & ER.3332cc & ER.333N1K & SBM0701 & SBM0901 \\
		\hline
		ER.333    & NA     & 0.01  & 0.13 & 0.07      & 0.01      & 0.13    & 0.13    \\
		ER.35     & NA     & NA    & 0.11 & 0.07      & 0.01      & 0.14    & 0.14    \\
		ER.5      & NA     & NA    & NA   & 0.16      & 0.13      & 0.24    & 0.24    \\
		ER.3332cc & NA     & NA    & NA   & NA        & 0.06      & 0.12    & 0.12    \\
		ER.333N1K & NA     & NA    & NA   & NA        & NA        & 0.13    & 0.13    \\
		SBM0701   & NA     & NA    & NA   & NA        & NA        & NA      & 0.00    \\
		SBM0901   & NA     & NA    & NA   & NA        & NA        & NA      & NA    \\
		\hline 
	\end{tabular}
\end{table}

\begin{table}[]
	\centering
	\caption{Wasserstein distances, real-world graphs} \label{Wd-r-w}
	\begin{tabular}{|l|cccccccc|}
		\hline
		& 1997/11/08 & 1997/11/09 & 1998/11/08 & 1998/11/09 & 1999/11/08 & 1999/11/09 & 2000/01/01 & 2000/01/02 \\
		\hline
		1997/11/08 & NA         & 0.00       & 0.03       & 0.03       & 0.04       & 0.04       & 0.04       & 0.04       \\
		1997/11/09 & NA         & NA         & 0.03       & 0.03       & 0.04       & 0.04       & 0.04       & 0.04       \\
		1998/11/08 & NA         & NA         & NA         & 0.00       & 0.02       & 0.03       & 0.03       & 0.02       \\
		1998/11/09 & NA         & NA         & NA         & NA         & 0.02       & 0.03       & 0.03       & 0.02       \\
		1999/11/08 & NA         & NA         & NA         & NA         & NA         & 0.02       & 0.02       & 0.01       \\
		1999/11/09 & NA         & NA         & NA         & NA         & NA         & NA         & 0.01       & 0.02       \\
		2000/01/01 & NA         & NA         & NA         & NA         & NA         & NA         & NA         & 0.02       \\
		2000/01/02 & NA         & NA         & NA         & NA         & NA         & NA         & NA         & NA        \\
		\hline
	\end{tabular}
\end{table}

\begin{table}[]
	\centering
	\caption{K-S distances and (p-values), synthetic graphs} \label{K-S-synth}
	\begin{tabular}{|l|ccccccc|}
		\hline
		& ER.333 & ER.35 & ER.5 & ER.3332cc & ER.333N1K & SBM0701 & SBM0901 \\
		\hline
		ER.333    & NA     & 0.43  & 1.00 & 0.15      & 0.11      & 1.00    & 1.00    \\
				&	& (0.00) & (0.00) & (0.00) & (0.00) & (0.00) & (0.00) \\
		\hline
		ER.35     & NA     & NA    & 1.00 & 0.46      & 0.38      & 1.00    & 1.00    \\
		& & & (0.00) & (0.00) & (0.00) & (0.00) & (0.00) \\
		\hline
		ER.5      & NA     & NA    & NA   & 1.00      & 1.00      & 1.00    & 1.00    \\
		&&&& (0.00) & (0.00) & (0.00) & (0.00) \\
		\hline
		ER.3332cc & NA     & NA    & NA   & NA        & 0.15      & 0.85    & 0.85    \\
		&&&& & (0.00) & (0.00) & (0.00) \\
		\hline
		ER.333N1K & NA     & NA    & NA   & NA        & NA        & 1.00    & 1.00    \\
		&&&&& & (0.00) & (0.00) \\
		\hline
		SBM0701   & NA     & NA    & NA   & NA        & NA        & NA      & 0.07    \\
		&&&&&& & (0.00) \\
		\hline
		SBM0901   & NA     & NA    & NA   & NA        & NA        & NA      & NA     \\
		&&&&&&& \\
		\hline
	\end{tabular}
\end{table}

\begin{table}[]
	\centering
	\caption{K-S distances and (p-values), real-world graphs} \label{K-S-r-w}
	\begin{tabular}{|l|cccccccc|}
		\hline
		& 1997/11/08 & 1997/11/09 & 1998/11/08 & 1998/11/09 & 1999/11/08 & 1999/11/09 & 2000/01/01 & 2000/01/02 \\
		\hline
		1997/11/08 & NA         & 0.00       & 0.01       & 0.01       & 0.01       & 0.02       & 0.02       & 0.01       \\
		&& (0.98) & (0.00) & (0.00) & (0.00) & (0.00) & (0.00) & (0.00) \\
		\hline
		1997/11/09 & NA         & NA         & 0.01       & 0.01       & 0.01       & 0.02       & 0.02       & 0.01       \\
		&&& (0.00) & (0.00) & (0.00) & (0.00) & (0.00) & (0.00) \\
		\hline
		1998/11/08 & NA         & NA         & NA         & 0.00       & 0.01       & 0.01       & 0.02       & 0.01       \\
		&&&& (0.87) & (0.00) & (0.00) & (0.00) & (0.00) \\
		\hline
		1998/11/09 & NA         & NA         & NA         & NA         & 0.01       & 0.01       & 0.02       & 0.01       \\
		&&&&& (0.00) & (0.00) & (0.00) & (0.00) \\
		\hline
		1999/11/08 & NA         & NA         & NA         & NA         & NA         & 0.01       & 0.02       & 0.00       \\
		&&&&&& (0.00) & (0.00) & (0.00) \\
		\hline
		1999/11/09 & NA         & NA         & NA         & NA         & NA         & NA         & 0.00       & 0.01       \\
		&&&&&&& (0.00) & (0.00) \\
		\hline
		2000/01/01 & NA         & NA         & NA         & NA         & NA         & NA         & NA         & 0.02       \\
		&&&&&&&& (0.00)\\
		\hline
		2000/01/02 & NA         & NA         & NA         & NA         & NA         & NA         & NA         & NA        \\
		&&&&&&&& \\
		\hline
	\end{tabular}
\end{table}

Our results show that our technique based on a transformation from graph to probability distribution and distance measurements with either the Wasserstein and K-S distances between node-node distributions are valid measures of network (dis)similarity. Indeed, these metrics accurately identify network structure changes, even in arguably difficult cases. For example, both metrics accurately detect the disconnection into two connected components of the ER graph with probability $p=0.333$ (ER.333 vs. ER.3332cc). 

Our results also confirm that our procedure correctly identifies the structural stability of the internet networks. In fact, our procedure is robust to degree outliers that are very common in real-world networks. Arguably, while the number of nodes and edges of internet networks do vary, the graph's connectivity structure remains constant. This robustness is reflected in the small distance between the distributions. 

Even so, here, we must also acknowledge the limitations of our method. While our network comparison technique is indeed robust to degree outliers, it does correctly detect changes in the number of edges and vertices and classify these networks as significantly different. However, the magnitude of their difference is very small, which is why we highlight robustness. For example, in the comparison between internet networks, our technique correctly identifies the difference between graphs 2000/01/01 and 2000/01/02 as statistically significant ($p=0.00$). As mentioned earlier, the magnitude of the difference is very low ($D,\tilde{\mathcal{W}} = 0.02$), in spite of a very significant difference in the number of edges and vertices. This low distance variation in response to large node and edge count variations in the distances between CDFs may be considered a limitation. For this reason, we caution against interpreting absolute magnitudes of distances without testing for significance.

Nevertheless, it must be noted that these graphs appear to be rather similar, from a structural point of view.  Indeed, both networks, have equal density and a very similar degree distribution. We posit that the magnitude of the difference remains small, although statistically significant, due to the structural similarity of these networks. Meanwhile, in the comparison between the ER graphs with 1,000 and 2,500 nodes (ER.333 vs. ER.333N1K), our technique did correctly identify a variation in network structure and a greater dissimilarity (distance) between these graphs. While these two graphs share the same edge probability parameter, their degree distributions differ significantly.

Finally, we must also offer a comparison of our technique to arguably simpler to obtain network characteristics, namely density and degree distribution. While density does indeed offer valuable information about a graph's structure, a comparison of densities is not sufficient to detect a change in structure. For example, the graphs ER.333 and ER.333N1K have identical densities, yet have significantly different degree distributions. Also, a graph's density does not offer any information regarding local connection patterns, such as community structure for example. 

Degree distribution also offers very valuable information about a network. Again, a comparison of degree distributions only offers a partial assessment of (dis)similarity. For example, the ER.333 and ER.333N1K graphs have significantly different degree distributions, yet have very similar connectivity patterns. Our two stochastic block model graphs (SBM0701 and SBM0901) have degree distributions that are very similar to the ER.333N1K, yet their connectivity (community) structure is totally different.

\section{Conclusion}
In this article, we present a statistical graph comparison technique which is based on node-node distances. Our results show that our technique accurately detects differences in graph structure. Future work will focus on statistical comparisons via sampling, which should offer greater scalability for our comparison technique. Naturally, we will also conduct further tests, using different scenarios.

\bibliography{biblio}

\end{document}